\documentstyle[twocolumn,prc,aps,epsfig]{revtex}
\input epsf
\newcommand{\be}{\begin{eqnarray}}
\newcommand{\ee}{\end{eqnarray}}
\def\lsim{\mathrel{\rlap{\lower3pt\hbox{\hskip1pt$\sim$}}
     \raise1pt\hbox{$<$}}} 
\def\gsim{\mathrel{\rlap{\lower3pt\hbox{\hskip1pt$\sim$}}
     \raise1pt\hbox{$>$}}} 

\begin{document}

\twocolumn[\hsize\textwidth\columnwidth\hsize\csname
@twocolumnfalse\endcsname

\title{What do lattice baryonic
  susceptibilities  tell us\\ about quarks,  diquarks and baryons at $T>Tc$?}

\author {Jinfeng Liao and
Edward V. Shuryak  }
\address { Department of Physics and Astronomy\\
State University of New York,
     Stony Brook, NY 11794-3800}

\date{\today}
\maketitle
\begin{abstract}
Lattice data on QCD thermodynamics, especially recent study of
high order susceptibilities by UK-Bielefeld collaboration, have
provided valuable information about matter properties around and above the
critical temperature $T_c$. In this work we tried to understand
what physical picture would explain these numerical data. We found
two scenarios which will do it: (i) a quark quasiparticle gas,
with the effective mass which is strongly $decreasing$ near the
phase boundary into the QGP phase; or (ii) a picture including
baryons at $T>T_c$, with the mass rapidly $increasing$  across the
phase boundary toward QGP. We further provide several arguments in
favor of the latter scenario, one of which is a natural continuity
with the baryon gas picture at $T<T_c$.
\end{abstract}
\vspace{0.1in}
]
\begin{narrowtext}
\newpage

\section{Introduction}

  QGP is experimentally studied via heavy ion collisions, at CERN SPS
and last years at BNL RHIC collider, at temperatures
reaching up to about $T\approx 2T_c$. Success of hydrodynamical
description
\cite{hydro} of observed collective flows have indicated, that
all dissipative lengths are very short and thus the produced
matter cannot be a weakly coupled gas but rather a near-perfect
(small viscosity) liquid
\cite{Shu_liquid}. These features are further complemented
by very high jet losses and robust
heavy quark charm (equilibration) observed, well beyond what pQCD
predicted.
As a result,
 a radically new picture of QGP at such temperatures
 is being developed, known as the {\em strongly coupled} quark-gluon
 plasma,
 or sQGP.

 It has been pointed
out by Shuryak and Zahed \cite{SZ_rethinking}
 that
the  interaction seems to be strong enough to preserve the
meson-like  bound states above $T_c$ although in a strongly
modified form. In particular, the lowest charmonium states
$J/\psi,\eta_c$ are predicted to exist up to $T$
as high as about $(2.5-3)T_c$.  These charmonium states were
observed on the lattice \cite{charmonium} as peaks in spectral
densities of the
correlation functions, and they indeed seem to survive till such high
temperatures.

It was further pointed out in the next paper by Shuryak and Zahed
 \cite{SZ_bound}
that in the deconfined phase
also  multiple binary {\em colored}
bound states should exist, in about the same $T$ domain,
since the interaction is about the same.
To put the discussion below
 into proper perspective, they argued
that there should be 3 categories of bound states,
in decreasing robustness: (i) glueballs, (ii)
$(qg)_3$
and mesons $\bar q q$; and (iii) $(qg)_6$, diquarks and
 baryons. If
 the strength
of the effective potential in $\bar q q$ states is counted as 1,
the relative color Casimirs for categories (i),(ii) and (iii) are
$9/4$, $9/8 \sim 1$ and $\approx 1/2$, respectively.
  In our recent work \cite{LS} we have extended the same approach
to some many-body states. We found new 3-gluon configuration $ggg$
belonging to category (i), the   polymeric chains $\bar q. g.g...g
q$ of the category (ii) and diquarks and baryons in category
(iii).

The last two are
the  baryon number carrying states we will discuss in this work.
Since
 these states belong to the $third$, most weakly
bound category, they are naturally most vulnerable to
uncertainties of the potential and their existence can be
questioned. Besides, these states are relatively heavy: such
states have not been included in
 \cite{SZ_bound} in pressure.

 The reason
we will discuss them now is because they are more important
at increasing
baryonic chemical potential $\mu$. Alternative way to
look at the same thing is to consider higher
derivatives over $\mu$ at $\mu=0$: this way the role
of such states is enhanced due
to powers of their baryon number. At some
point
the
 diquarks and baryons  should become noticeable in these quantities
even if their role in pressure is small: and
this  is precisely what we think happened
in the lattice data of the UK-Bielefeld collaboration (UKB)
 \cite{ci_paper}, especially in
  susceptibilities with 4 and 6 derivatives.

 In this work we will concentrate on the so called baryonic
susceptibilities
 part of the free energy, which can be
singled out via derivatives over quark chemical potentials
$\mu_q=(\mu_u+\mu_d)/2$ and $\mu_I=(\mu_u-\mu_d)/2$ calculated
recently   by the UKB. They use it in a context of Taylor
expansion of the
 thermodynamical quantities in powers of  baryonic chemical potential
 $\mu/T$\footnote{
We follow notations used in this work where $\mu$ is the chemical
potential
per quark, not
per baryon. Thus the associated charge is $B=1$ for a quark,  $B=2$
for a
diquark and $B=3$ for a baryon.
}
up to the order $O(\mu^6)$ of 2-flavor QCD,
but we will not discuss this expansion per se and concentrate on
(T-dependent) susceptibilities of the kind\footnote{Since we would not discuss any Taylor series in this work,
we would prefer to leave out the factorials and thus discuss
susceptibilities $d_n$ defined without them, not $c_n$.}:
\be d_n(T)= {\partial^n (p/T^4)\over \partial
  (\mu/T)^n}{\bigg | }_{\mu=0} = n !c_n(T)
\ee for $n=2,4,6$. (The odd ones vanish at $\mu=0$ by symmetry.)
These data are shown in Fig.\ref{fig_baryonfit} and also below.
The UKB also studied what they called isospin susceptibilities
defined as \be d^I_n(T)= {\partial^n (p/T^4)\over \partial
  (\mu/T)^{n-2} \partial (\mu_I/T)^2}{\bigg | }_{\mu=\mu_I=0} = n
  !c^I_n(T) \ee and in a flavor diagonal-non-diagonal language there are
\be d^{uu}_n=(d_n+d^I_n)/4 \, , \, d^{ud}_n=(d_n-d^I_n)/4 \ . \ee
Let us also mention another recent independent lattice studies on
 susceptibilities of  2-flavor QCD in \cite{gupta}, where they
 defined so-called nonlinear susceptibilities(NLS)
\be \chi_{n_u,n_d}=\frac{\partial^{(n_u+n_d)} p}{\partial
\mu_u^{n_u}
\partial
 \mu_d^{n_d}} \ee and evaluated their values on lattice. To see the connection
 between these two approaches, we give the following relations
\be d^{uu}_n=T^{n-4}\sum_{l=0}^{n-2} C_{n-2}^l
 [\chi_{l+2,n-l-2}+\chi_{l,n-l}]/2 \, , \ee and \be d^{ud}_n=T^{n-4} \sum_{l=0}^{n-2} C_{n-2}^l
 \chi_{l+1,n-l-1} \, .\ee
 While the two approaches are very closely related, their numerical results, however,
 are not quantitatively comparable, partly because they have used very different mass setup in
 the lattice calculation. We nevertheless emphasize that in a qualitative view both of them have
 found very similar and interesting patterns in those susceptibilities, especially for the 4th and
 the 6th, which are the central issues to be addressed in this paper.

To set the stage, we start with the hadronic phase below $T_c$.
Here the relevant states are only the baryons with the baryon
number (per quark) 3. Their spectrum is known at $T=0$
experimentally, and thus an obvious question is: Can a simple
resonance gas of known baryons explain the behavior of these
susceptibilities below $T_c$? Indeed it is the case, as shown by
the dotted curves in Fig.\ref{fig_baryonfit} (obtained by
including contributions of nucleon states from N(940) to N(1675)
and $\Delta$ states from $\Delta$(1232) to $\Delta$(1700),
 for two-flavor theory one should not include strange
baryons). No $T-$ or $\mu-$ dependence of these masses is assumed,
nor do we take into account the fact that lattice is dealing with
non-massive quarks\footnote{In fact
  the input quark mass in these calculations is $.4T$.}: tuning
  these will shift the curves down a bit, making the agreement even better.
So the susceptibilities in the hadronic phase,  $T<T_c$,
 can be described by the usual
resonance gas of baryons.


\begin{figure}
\hspace{+0cm} \centerline{\epsfxsize=7.5cm\epsffile{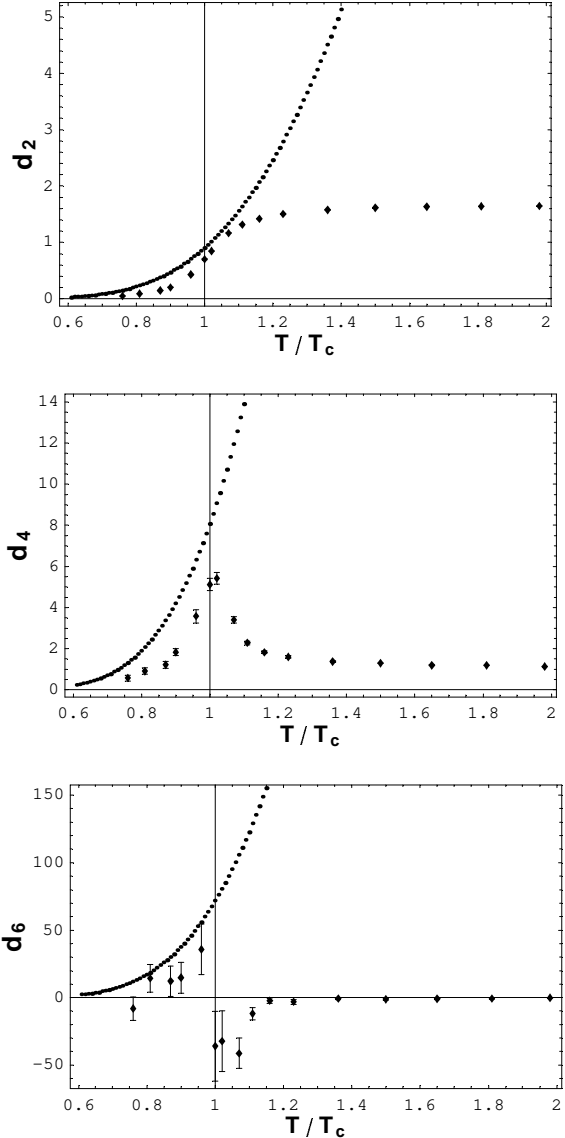}}
\hspace{+0.5cm}
 \caption{\label{fig_baryonfit}
The dotted lines correspond to a gas of baryonic resonances, the
points with error bars are the susceptibilities
$d_2(T),d_4(T),d_6(T)$ from lattice data (after removal of
factorials in $c_{2,4,6}$). }
 \end{figure}

The main issue to be discussed in this work is what these lattice
data actually tell us about the nature of baryonic states $above$
$T_c$, and whether one can describe them with sQGP model or as
well with some other model.

  Before we proceed to the argumentation in literature, let us
remind standard thermodynamical expressions for massive fermions,
which can be put in the following well known form
\begin{eqnarray}
\label{eq_p}
 {p\over T^4}= N {M^2\over 2 \pi^2
  T^2} \sum_{l=1}^{\infty}{\bigg [}{(-)^{l+1}\over l^2}
 && \, {\big ( } e^{lB\mu/T} \nonumber \\ &&+e^{-lB\mu/T}{\big )}K_2(lM/T){\bigg ]}
  \end{eqnarray}
where $B$, $M$ is the baryon number of the corresponding particle
and its mass, $N$ is the statistical weight and $K_2$ is the
Bessel function.\footnote{If there are more than one species of
particle we then sum over different species. Yet there will be
particular concern when dealing with quasiparticles instead of
particles where some background term may arise in the pressure, as
will be discussed in later section.
 } This form is very convenient for taking derivatives over
$\mu$, for example the first derivative, the baryon density is
\begin{eqnarray}
\label{eq_n}
 {n_B\over T^3}=&& N B {M^2\over 2 \pi^2
  T^2} \sum_{l=1}^{\infty} {\bigg [ } {(-)^{l+1}\over l}  \,
 {\big (} e^{lB\mu/T} \nonumber \\ && \qquad \qquad \qquad \qquad \qquad
 -e^{-lB\mu/T}) K_2(lM/T) {\bigg ]} \nonumber \\
 =&& N B \,\,{\cal N}[B\mu/T,M/T]
 \end{eqnarray}
where the function ${\cal N}[x,y]$ is defined by these series.
Note at this point we don't really consider mass as depending on
$\mu$ so no extra derivatives against $M$ appear.

In a number of talks Karsch (and also a  recent preprint)
\cite{Karsch_talks} have presented what we would refer to as a
``naive'' argument:  the subsequent ratios
\be  \label{eqn_B**2} d_{n+2}/d_n\approx  <B^2> \ee
 are
directly related to the squared baryon number of the constituents. The
argument goes as follows: (a) For massive particles with $M>>T$
one can use the so called Boltzmann approximation, keeping only
the first term in the sum above; (b)  after that
 the $\mu-$ dependence factorizes, and thus
each two derivatives over $\mu$ restore the same expression, modulo
the factor $B^2$.
In the matter dominated by quark quasiparticles, or $qg$ bound states,
the r.h.s. would be 1, but it would instead be 4 or 9 for
matter dominated by diquarks or baryons, respectively.
 The measured ratio $d_4/d_2$
is $\sim 10$ at $T<T_c$ but at $T>T_c$ it rapidly drops  and becomes
close to 1. Comparing it to the formula above  Karsch concluded
that at $T>T_c$  matter is a gas of some $B=1$ objects,
 while the contribution of the  $B=2$ diquarks
is strongly restricted.

 But if one looks closer at this argument, one  finds
 it missing a lot of effects that should be there as
well. For example, the next similar ratio $d_6/d_4$ above $T_c$
is nowhere close to 1
 but is in fact a large negative number $\sim -10$ which cannot be
interpreted as a $B^2$ of anything.

 Furthermore, the  idea that one can keep only the
main term in the sum so that the $\mu$ and $T$ dependence can
factorize, must be wrong by itself. The $T$-dependence of
$d_2(T)$, $d_4(T)$ and  $d_6(T)$ is not at all similar: while
$d_2(T)$ resembles the behavior of the pressure itself and can
easily be interpreted as a transition from hadron to quark gas,
the next one $d_4(T)$ has a sharp maximum near $T_c$, with even
more complicated ``wiggle'' in the  $d_6(T)$.

Another perspective on that issue can be made if one converts
baryon number and isospin susceptibilities into flavor-diagonal
($uu$ or $dd$) and flavor non-diagonal   $ud$ susceptibilities.
The lattice data show that the second flavor-mixing  derivatives
are small\footnote{This is also the main point of the paper
\protect\cite{Baryon_Strangeness}.} $d_2^{ud}/d_2^{uu}<<1$, but
similar ratios for higher derivatives n=4,6 are not small
$d_n^{ud}/d_n^{uu}\sim 1/2$.

Does it imply that the quark gas model is also inadequate and
should be
 excluded as well as the ``bound state'' gas?
Or, if the argument is wrong, what exactly is missing?
\\
(i) Even if the Boltzmann approximation (keeping the first term in
sum in (\ref{eq_p})) may be good for pressure, it still fails
for higher susceptibilities because the $l$-th term has $l\mu$ in
the exponent, and subsequent differentiation their role grows as
$l^n$. By the time one comes to the sixth derivative, these terms
start canceling each other. In physics terms, this is a form of
Fermi blocking effect not included in the simple Boltzmann
approximation.

(ii) The second to recognize is the fact that quasiparticles are
not particles and their effective masses depend on matter
parameters, such as $T$ and especially $\mu$. Subsequent
differentiation of this effective mass over $\mu$ would add powers
of derivatives like \be M''=\frac{\partial^2 M}{\partial \mu^2}
(T,\mu=0) \ee to susceptibilities and to their ratios such as
(\ref{eqn_B**2}). Provided those are large enough, they may
completely invalidate the naive interpretation of those ratios as
baryon number squared. This was already pointed out by Bluhm,
Kempfer and Soff  \cite{BKS}, and we will refer to it below as the
``BKS effect''.  The same is true for bound states
such as baryons, and similar derivatives of their masses  $M''_B(T)$
would play an important role below.
\\
(iii) the contribution of diquarks has been grossly overestimated, while
the contribution of baryons was not discussed at all. We will
 show below that it may naturally explain the features seen in higher derivatives.
 \\

  The outline of the paper is as follows. In section \ref{sec_unconstrained}
we will start with an ``unconstrained'' quark gas model, and will
use the lattice data to extract
 the quasiparticle mass  together
with its dependence on matter, $M(T,\mu)$. We would not need to rely
on perturbative arguments used by BKS \cite{BKS} (since even their own
fit leads to rather strong coupling at $T\approx T_c$).
Furthermore, we will conjecture  possible relation between the
 $T-$ and $\mu-$ dependences due to known shape
of the phase boundary on the phase diagram. In section
\ref{sec_constrained} we will further impose a number of
constraints on quark mass, from other lattice data and also from
confinement, a condition that there should not be any colored
degrees of freedom at $T<T_c$. We will conclude that these
constraints basically make it impossible to ascribe  the observed
features of the data  to the BKS effect. After that we will proceed to section
\ref{sec_bar} in which we will discuss the contribution of diquark
and baryons: here we will find good fits to the data satisfying
all the needed constraints and nicely joining the baryon gas
picture below $T_c$.

\section{Model I: a quark gas with an unconstrained  mass $M(T,\mu)$ }
\label{sec_unconstrained}
  The idea to use thermodynamical quantities calculated on the lattice
to fit the mass parameters of quasiparticles is by itself quite old. For example,
Levai and Heinz \cite{Levai:1997yx} have used the data on $p(T)$ for
determination of quark and gluon effective masses $M(T)$\footnote{It was not
as
direct as our approach below,
because one cannot get 2 functions out of one without assumptions.}.

One well known problem with quasiparticle gas models is that
the derivatives over $T$ and $\mu$ upset thermodynamical consistency
between
gas-like expressions for different thermodynamical quantities.
Only one of them can  be assumed to have a
simple additive form over  quasiparticles: then there is
no freedom left and all other quantities can be calculated
from it by thermodynamics. Thus only one ``primary''
expression can be additive, while others will have extra
``derivative'' terms complementing
 simple gas formulae.

Following conventions of the BKS paper, we will use as such
``primary'' expression that for the baryon number density
  (\ref{eq_n}).  The expressions
for pressure and energy density would then be corrected by some
$T,\mu$-dependent ``bag terms''. Higher derivatives terms $d_n$
will be calculated by differentiating (\ref{eq_n}) $n-1$ times. To
be more specific, we explicitly give the baryon number density for
this quark gas model
\begin{equation} \label{density}
\frac{n_B}{T^3}=\frac{\partial (p/T^4)}{\partial
(\mu/T)}=\frac{g}{2\pi^2}\int dx x^2 \, n\,[F(\epsilon-n
\tilde{\mu})-F(\epsilon+n \tilde{\mu})]
\end{equation}
Here $g=N_s*N_c*N_f=12$ is the degeneracy factor for quarks in the
two-flavor case, $n$ is the baryon quantum number of quark which
is defined here to be $n=1$ by setting $\mu$ to be the quark
chemical potential. $\tilde{\mu}=\mu/T$ is made to be
dimensionless, and $\epsilon=\sqrt{x^2+\tilde{m}}$ with
$\tilde{m}=M/T$. And finally we have introduced Fermi distribution
function $F(y)=\frac{1}{e^{y}+1}$. Starting from (\ref{density})
the explicit formulae for $d_2,d_4,d_6$ are given to be:
\begin{eqnarray}\label{c2}
d_2 = \frac{\partial (n_B / T^3)}{\partial \tilde{\mu} }{\bigg
|}_{\mu=0} = - \frac{2g}{ 2\pi^2 } \int dx x^2\, n^2
F^{(1)}(\epsilon_0) \qquad
\end{eqnarray}
\begin{eqnarray}\label{c4}
d_4 &=&  \frac{\partial^3 (n_B / T^3)}{\partial
\tilde{\mu}^3 }{\bigg |}_{\mu=0} \nonumber \\
&=& - \frac{2g}{ 2\pi^2 } \int dx x^2 \, {\bigg [ }n^4
F^{(3)}(\epsilon_0) \nonumber \\
&& \qquad \qquad + 3 n^2 \,
 F^{(2)} (\epsilon_0) \, \frac{\tilde{m}_0}{\epsilon_0} \,
{\big ( } \frac{\partial^2 \tilde{m}}{\partial \tilde{\mu}^2}{\big
|}_{\mu=0} {\big ) } {\bigg ] } \qquad \quad
\end{eqnarray}
\begin{eqnarray}\label{c6}
d_6 &=&  \frac{\partial^5 (n_B / T^3)}{\partial
\tilde{\mu}^5 }{\bigg |}_{\mu=0} \nonumber  \\
&=& - \frac{2g}{ 2\pi^2 } \int dx x^2 \, {\bigg [ }n^6
F^{(5)}(\epsilon_0)  \nonumber \\
&& \qquad + 10 n^4 \,
 F^{(4)} (\epsilon_0) \, \frac{\tilde{m}_0}{\epsilon_0} \,
{\big ( } \frac{\partial^2 \tilde{m}}{\partial \tilde{\mu}^2}{\big
|}_{\mu=0} {\big ) } \nonumber \\
&&\qquad + 15 n^2 \,
 F^{(3)} (\epsilon_0) \, \frac{\tilde{m}_0^2}{\epsilon_0^2} \,
{\big ( } \frac{\partial^2 \tilde{m}}{\partial \tilde{\mu}^2}{\big
|}_{\mu=0} {\big ) }^2 \nonumber \\
&&\qquad + 5 n^2 \,
 F^{(2)} (\epsilon_0) \, {\bigg ( } \frac{\tilde{m}_0}{\epsilon_0} \,
{\big ( } \frac{\partial^4 \tilde{m}}{\partial \tilde{\mu}^4}{\big
|}_{\mu=0} {\big ) } \nonumber \\
&&\qquad + \frac{3x^2}{x^2+\tilde{m_0}^2} {\big ( }
\frac{\partial^2 \tilde{m}}{\partial \tilde{\mu}^2}{\big
|}_{\mu=0} {\big ) }^2 {\bigg ) } {\bigg ] } \qquad \qquad \qquad
\end{eqnarray}
In the above equations we have used
$\epsilon_0=\sqrt{x^2+\tilde{m_0}^2}$ and
$\tilde{m}_0(T)=M(T,\mu=0)/T$, and also $F^{(i)}(y)$ means the
$i$th derivative of the function $F(y)$.

The model used in the BKS paper assumes
 some Hard Thermal Loop based perturbative form for the $T,\mu$-dependent mass
 with the  coupling $g^2(T,\mu)$ running in a complicated fashion
 fitted to reproduce
the  susceptibilities we discuss in this work. However, we do not
see why any assumptions about the mass dependence are actually
needed\footnote{There is of course no reason to trust any
perturbative formula  near $T_c$ at all, where the coupling
becomes as strong as it was found by BKS themselves. } at this
point.

We thus suggest a generalization of what was done in \cite{BKS}.
Assuming a simple ideal gas model of  quark quasiparticles, one
has their mass to be the only input needed.
   With the lattice data on  $d_2(T)$, $d_4(T)$ and $d_6(T)$
  used as input,
one can simply solve for the three
 functions of $T$ which would ideally fit them: we have chosen those to be:
(i) the quark mass
$M(T,\mu=0)$; and its two lowest non-zero\footnote{The quasiparticle
   masses
and other quantities obviously can depends only quadratically on
$\mu$ because of $\mu\rightarrow -\mu$ symmetry based on CP
invariance. } derivatives over $\mu$ (ii) $M''=\frac{\partial^2
M}{\partial \mu^2} (T,\mu=0)$ and (iii) $M''''=\frac{\partial^4
M}{\partial \mu^4} (T,\mu=0)$. With these at hand, of course, we
are able to develop the Taylor's expansion for quark mass as a
function of $|\frac{\mu}{T}|<1$:
\begin{eqnarray} \label{mass_expansion}
M(T,\frac{\mu}{T})=M(T,\mu=0)&&+ \frac{1}{2!}\frac{\partial^2
M}{\partial \mu^2} (T,\mu=0)\cdot(\frac{\mu}{T})^2 \nonumber
\\ &&+\frac{1}{4!}\frac{\partial^4 M}{\partial \mu^4}
(T,\mu=0)(\frac{\mu}{T})^4
\end{eqnarray}

\begin{figure}
\hspace{+0cm} \centerline{\epsfxsize=8.cm\epsffile{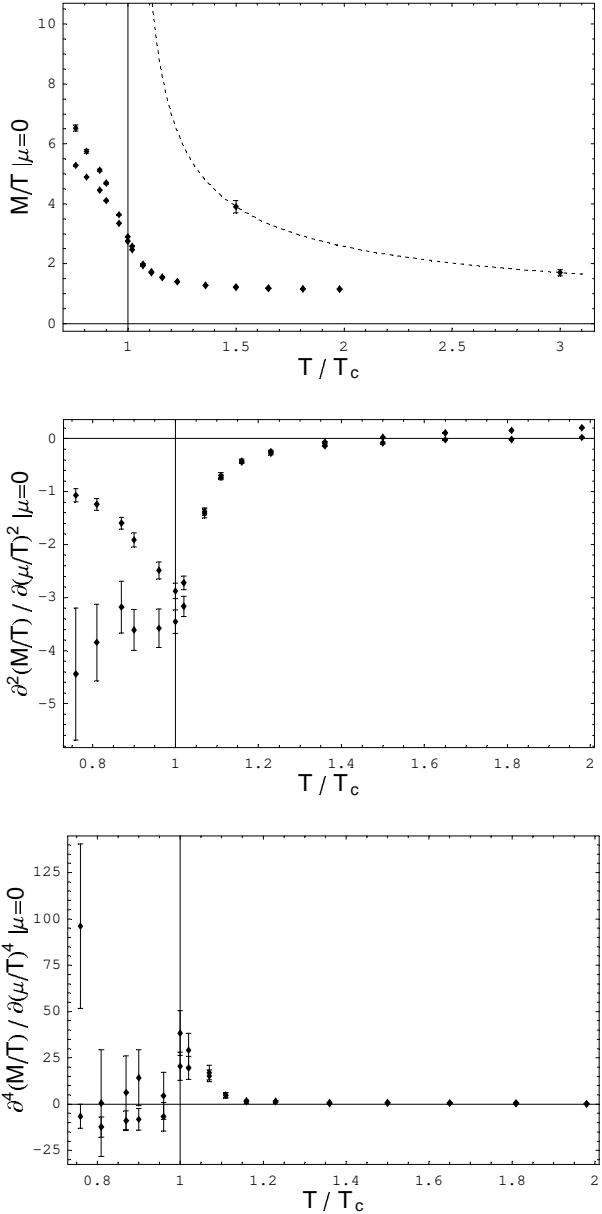}}
\hspace{+0.5cm}
 \caption{\label{fig_mass_c_ci}
 Quark quasiparticle mass and its second and fourth
 derivatives over $\mu$ as a function of temperature $T$, extracted from lattice
 data for susceptibilities. There are two sets of points in each figure that are obtained from
 $c_2,c_4,c_6$ and from $c^I_2,c^I_4,c^I_6$ respectively. In the
 top figure for quark mass, we also plotted the two points with error bars measured by lattice via propagator, and
  the mass given by (\ref{quark_mass}) as well.  (the dashed line).
}
 \end{figure}

The procedure is iterative: First we used $c_2(T)$ data to solve
for the mass $\tilde{m_0}$ as unknown. Then we go to $c_4$, the
equation of which includes both $\tilde{m_0}$ and
$\frac{\partial^2 \tilde{m}}{\partial \tilde{\mu}^2}{\big
|}_{\mu=0} $, but since we have already solved $\tilde{m_0}$ from
$c_2$ now the only unknown term is $\frac{\partial^2
\tilde{m}}{\partial \tilde{\mu}^2}{\big |}_{\mu=0} $, which could
be solved out from lattice results of $c_4$. finally we can obtain
$\frac{\partial^4 \tilde{m}}{\partial \tilde{\mu}^4}{\big
|}_{\mu=0} $ from $c_6$ with $\tilde{m}_0$ and $\frac{\partial^2
\tilde{m}}{\partial \tilde{\mu}^2}{\big |}_{\mu=0} $ already being
solved from $c_2$ and $c_4$. The results for these three functions
are shown in Fig.\ref{fig_mass_c_ci}. (The error bars in
$\tilde{m}_0$ are determined from uncertainty in $c_2$. While for
$\frac{\partial^2 \tilde{m}}{\partial \tilde{\mu}^2}{\big
|}_{\mu=0} $ the errors should come from both $c_4$ and
$\tilde{m}_0$, the error bars in the figure only include those
from $c_4$, and also for $\frac{\partial^4 \tilde{m}}{\partial
\tilde{\mu}^4}{\big |}_{\mu=0} $ the error bars solely include
that originated from $c_6$.)

As an independent check, we have also extracted the same three
quantities, $M(T,\mu=0)$, $\frac{\partial^2 M}{\partial \mu^2}
(T,\mu=0)$ and $\frac{\partial^4 M}{\partial \mu^4} (T,\mu=0)$
from the lattice data set for $c^I_2(T)$, $c^I_4(T)$ and
$c^I_6(T)$ from \cite{ci_paper}  by the same strategy (but
starting with isospin densities).

  The results are shown in Fig.\ref{fig_mass_c_ci}. As
can be seen, two sets of parameters we extracted from both data sets
 are well consistent with each other at $T>T_c$,
while for $T<T_c$  they  do not agree. It is a good feature, as
 the quark gas model is not supposed to work there, in the domain of the
 baryon
resonance gas.

Let us summarize these results. The most important
 lessons are: (i) the mass $M(T)$  strongly increases when cooling down toward
the critical point $T_c$;
(ii) Large and  $negative$ $\frac{\partial^2 M}{\partial \mu^2}
 (T,\mu=0)$ close to
 $T_c$;  (iii) The
 4-th derivative is positive: so this decrease of the mass due to the
 2nd derivative will stop at about $\mu/T\sim 1$, see (\ref{mass_expansion}).

  The first two points are the trends already emphasized
 by BKS \cite{BKS}. In their approach these two features are
related  with each other because of the assumed perturbative
origin of the effective quark mass: \be \label{perturbative}
M=g^2(T,\mu) T^2\left(1+N_f/6+{1  \over 2  \pi^2 T^2}
\sum_f\mu_f^2\right)\ee where the sum runs over all flavors $f$.
 Ignoring for a moment a (rather complicated)
running of the coupling,  the BKS mass is thus constant at the
particular ellipsoids in the $T-\mu$ plane, thus the derivatives
over $T$ and $\mu$ are related.

We would like to propose another reasoning that leads to similar
effect, but is free from perturbative assumptions. Its idea can be
described as follows: the quark mass should be getting large not
only near the critical point $T\rightarrow T_c,\mu=0$, but near
the {\em whole critical line}, at all $\mu$. It is needed
 to ensure that quark degrees of freedom do
not contribute in the confined phase, at any $\mu$.

The  critical line at nonzero $\mu$ is schematically shown in
Fig.\ref{fig_radial_tmu}, its shape at not-too-large $\mu/T$ can
be described by an ellipsoid, or an unit circle, if the units are
chosen appropriately. One may further think that the mass
dependence on the radial coordinate $R$ on such a plot is much
more important than on the angular one $\phi$ since  the ``lines
of constant mass'' should  be nearly parallel to the critical
line, at least in its vicinity where the discussed effect takes
place.\\

\begin{figure}
\centerline{\epsfxsize=8.5cm\epsffile{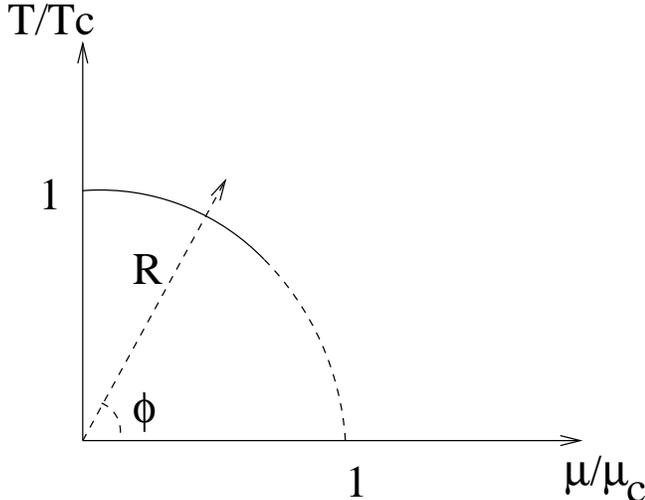}}
\caption{\label{fig_radial_tmu}
 In the plane of temperature $T$-baryonic chemical
potential $\mu$, both appropriately normalized, the phase boundary
looks like a part of a circle. (At least for the part marked by
the solid line, studied well at SPS and RHIC, with quite well
established chemical freezout. The dashed line is a continuation
of the freezout line where its association with the critical line
is questionable.)
 The polar coordinates to be used are
the radial distance $R$ and the angle $\phi$.}
 \end{figure}

So, the proposed extension of the $T$-dependence of the mass to
its $\mu$-dependence is based on a substitution \be
\label{mass_mu}  M(T,\mu=0)\rightarrow M(R(T,\mu)) \ee \be R^2={T^2
\over T_c^2}+{\mu^2 \over \mu_c^2} \ee We have  introduced here a
new parameter $\mu_c$: its value can be readily obtained from the
experimental freezeout curve measured in heavy ion collisions at
small $\mu$, believed  to represent the critical line. If so,
 the value of this parameter is
\be \mu_c/T_c=1.7 \ee which is quite different from the value
given by ``perturbative scaling'' (\ref{perturbative}):
$\sqrt{(1+{N_f \over 6})\pi^2}\approx 3.63$ which is not supposed
to work in the non-perturbative regime near $T_c$.

\section{Model II: the constrained quark gas }
\label{sec_constrained}
  The ``unconstrained Model I'' discussed above, although consistent with both data
  sets
$d_n(T), d^I_n(T)$, is
 unfortunately unacceptable, for two main  reasons:
(i) It  contradicts direct lattice measurements of the
quasiparticle masses ; (ii) It implies that quark degrees of
freedom still significantly contribute in the confining phase at
$T<T_c$.
In this section we will show what happens if one tries to modify
the unconstrained model to make it compatible with both.

One feature of the Model I is the relatively light quark mass
$M(T,\mu=0)$ in region $1-2 T_c$  ranging from about $1.7T_c$ to
$2.2T_c$. Such  mass conflicts with another lattice data about
quark quasiparticle mass at $1.5T_c$ and $3T_c$, see \cite{masses}
which are
 $m_q/T=3.9\pm 0.2$ and  $m_q/T=1.7\pm 0.1$, respectively,
and are
shown in Fig.2 by two crosses with the error bars.
Although these results are based on only one paper and have not been
systematically studied by other lattice groups so far, they  nevertheless
represent direct measurements from the quark propagators.
 Furthermore, such large
masses  correspond to the inter-particle potentials at large
distances measured in separate lattice study \cite{THERMO}.

Although the mass extracted via the Model I grows toward $T_c$,
this effect is still not robust enough to make quark contribution
near-zero (or negligible) at $T=T_c$. (In fact, BKS proceeded to
fit equally well some region below $T_c$.) This is unacceptable,
since we know that there are no propagating quark degrees of
freedom in the confining phase.

Both these reasons force us to reconsider Model I, basically by
increasing the quark mass significantly to meet both constraints.
This can be achieved by a quark mass formula similar to that used
in \cite{SZ_bound}
\begin{equation} \label{quark_mass}
M(T)={0.9 \over T-1} + 3.45+0.4T
\end{equation}
with all units in proper powers of $T_c$ (This and subsequent
mass formula would then be generalized to finite $\mu$ according
to (\ref{mass_mu})
. The coefficients are chosen
so that the curve goes through the two lattice-measured points
for quark mass at $T=1.5,3.0\,T_c$, see the dashed line in
Fig.\ref{fig_mass_c_ci}.

We show what happens then to the susceptibilities, see the
medium-thickness solid lines in Fig.\ref{fig_c4_baryons}. In
short, good description of $c_2(T)$ is definitely ruined The issue
is the same as for pressure in \cite{SZ_bound}, and perhaps can be
cured by $qg$ and other bound states. But this is not the only
problem of the constrained model: although it can produce a peak
in $d_4(T)$ and a ``wiggle'' in $ d_6(T)$, given large enough
derivatives over $\mu$, those get displaced toward larger $T$ as
compared to the data. It is an inevitable consequence of the
second constraint, insisting that quark effect be effectively zero
at $T_c$.\footnote{The very heavy mass due to the constraints
significantly decrease quark contribution to thermodynamics and
hence disfavor quark-only model, yet on the other hand, it
strongly favors the formation of bound states.}

Completing our discussion of purely quark models, we now proceed
to the possible role of their bound states, diquarks and baryons.

\section{The effect of diquarks and baryons}
\label{sec_bar}
  We will now proceed to contributions of the bound states to the
  baryonic susceptibilities. Let us remind the reader that the
 particular reason to focus on diquarks and especially baryons is
that the role of diquarks and baryons relative to quarks grows
with $\mu$ because of their larger baryon charges. Alternatively,
their contribution to the susceptibility $d_n$ grows exponentially
 with $n$:  by the factors $2^n$ for  the diquarks and and $3^n$ for baryons. For
example, the contribution of $N,\Delta$ is enhanced by a factor 81
for $d_4$ and 729 for $d_6$ relative to pressure: with estimates
of pressure given above one may then expect to see their
contribution there.
 On the other hand,
for lower derivative $d_2$ we expect quark-gluon bound states,
which are more numerous and more tightly bound, to contribute
significantly. We summarized all bound states, together with their
multiplicities,  in the
Table.\ref{table_states}.

(In passing, let us comment about the numbers in the real world
with strangeness, $N_f=3$. The number of diquark flavor states is
 increased to be 3 times larger, for baryons the total spin-flavor multiplicity increases
from 4+16=20 to 56(an octet $J=1/2$ and a decuplet $J=3/2$) which
is roughly enhanced by 3 times, so the numbers both diquarks and
baryons states are increased by the factor 3. The quark number
increases as 3/2, so the overall enhancement of the ratios we will
discuss below from $N_f=2$ to $N_f=3$ is the factor 2.)

\begin{table}
\caption{\label{table_states} Summary of states with baryon number
at $T>T_c$ studied in this paper. }
\begin{tabular}{ccccc}
state  & spin & flavor & color & multiplicity  \\
$q$ &      2 & 2 & 3  & 12  \\
$(qg)_3$ & 4 & 2 & 3 & 24  \\
$(qg)_6$ & 4 & 2 & 6 & 48  \\
$(qq)_3^{J=I=0}$  & 1 & 1 & 3  & 3 \\
$(qq)_3^{J=I=1}$  & 3 & 3 & 3  & 27 \\
$N$ & 2 & 2 & 1 & 4 \\
$\Delta$ & 4 & 4 & 1  & 16 \\
\end{tabular}
\end{table}

{\bf Quark-gluon bound states}: Before we proceed to
actual calculation, let us make simple estimates of the relative
weighing in pressure of quark-gluon bound states. The 2-body
states $qg$ are thermodynamically suppressed by additional Boltzmann factor,
$exp(-M/T)\sim 0.02-0.04$ (by including their considerable
binding). However, due to their relatively large multiplicity (6
times the number of the quark states) they contribute to the
pressure and susceptibilities at the level of about 1/10 or more.

To get more quantitative answer one has to know
 the binding energy of these states.
While the binding of the category three states
$(qg)_6$ can be reasonably neglected, the category two
$(qg)_3$ states have considerable binding at the same order as
meson states. The potential model calculations in \cite{LS} lead
to $(qg)_3$ binding up to ${|\delta E| / T}\approx 1.4$ at
$T=T_c$, which means their contribution increases
relative to simple estimate above by extra factor 2-3.

  However there are many reason to doubt that close to $T_c$
this calculation can be trusted quantitatively. In particular,
 the potential used is
measured on the lattice for $static$ charges only, and the
corresponding calculations are supposed to be reliable only when
the binding is small: near $T_c$  more complicated
dynamics beyond the potential model will  contribute as well.

\begin{figure}
\hspace{+0cm}
\centerline{\epsfxsize=7.cm\epsffile{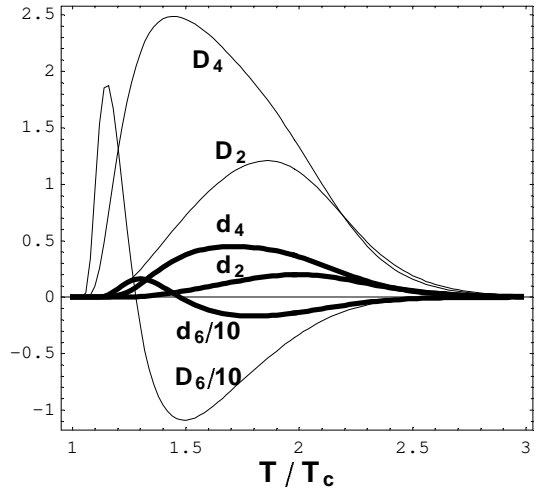}}
\hspace{+0.5cm}
 \caption{\label{fig_qg}
Comparison of susceptibilities from quark-gluon states with two
limiting case, zero-binding ($d_{2,4,6}$) and "full-compensation"
binding ($D_{2,4,6}$).}
 \end{figure}

Let us thus just suggest an {\em upper limit} for the $qg$ states'
contribution. Since the $qg$ states are colored,  they should gets
infinitely heavy at $T_c$, together with all other colored states.
Furthermore,
 as (the more tightly bound)
$qg_3$ states have the total charge of one quark, their mass
should not be smaller than that of one charge
$|\delta E|<M_q(T\approx T_c)$. So we expect $M(qg)$ to interpolate between
 $M_q+M_g\approx 2M_q$ at zero binding to a
 a single $M_q$ at $T\approx T_c)$.
The contribution of these states to susceptibilities in the
two limiting cases, namely the zero-binding case (labeled in
figure by $d_i$) and "full-compensation" binding case (labeled by
$D_i$), are shown in Fig.\ref{fig_qg}. We conclude that
large uncertainty, of the order of factor 3, remains in the
contributions
of such states.

 (These results are calculated with $(qg)_6$ always having twice
quark mass and melting at $1.4T_c$ while with $(qg)_3$ having
twice quark mass in the former case and the same mass as quark in
the latter, both melting at $2.1T_c$. The actual contribution of
quark-gluon bound states should be somewhere in between, near to
$D_i$ around $T_c$ while rapidly decreasing to $d_i$ for higher
temperature.)

 One may also ask what is the contribution of the
various polymer-like qg states $qg,\, qgg, \, qggg, \cdot \cdot
\cdot$ which, according to \cite{LS},
has
 the same
binding energy per bond. The effect of these states can be easily
evaluated via a geometric series: the resulting
 enhancement  factor is
\be \label{eqn_poly} f_{polymers}={1 \over
1-6 exp[(|\delta E|-M_g) / T] } \ee
where 6 is the color and spin degeneracy added by each link.
For small binding this is just a few percent correction, but if it
may get to be strong enough to drive the denominator toward zero,
a total ``polymerization'' of sQGP would occur.

 {\bf Diquarks}:
 for $N_f=2$ gauge theory corresponding to the UKB data at hand
there is only one attractive diquark channel, the antisymmetric
color triplet\footnote{ In the talks
 \cite{Karsch_talks} Karsch  mentioned ``hundreds of diquark bound states''
in our model: it must be some misunderstanding.}
 $(qq)_3$. Because of
Fermi statistics, it means  that the product of spin and flavor
should be symmetric, and thus there are two options: (i) spin-0
isospin-0 $ud$ diquark $(qq)_3^{J=I=0}$, and (ii) spin-1 isospin-1
one $(qq)_3^{J=I=1}$. These are the diquarks which are familiar in
hadronic spectroscopy, the former appears inside the $N$, the
latter inside $\Delta$ (octet and decuplet members, for 3
flavors). The lesson from this spectroscopy (at T=0, of course) is
that while the former is well bound, by about 300 MeV, the latter
is not. In view of the rather marginal character of diquark
binding, we expect only the former one able to be seriously
considered as bound state above $T_c$. Nevertheless to confirm the
point that diquarks will not play any role in all susceptibilities
measured, we include both of them in calculation of
Fig.\ref{fig_c4_baryons}. If we only use antisymmetric states,
then the contribution will be reduced to only 1/10 of that.

The diquark-to-quark pressure ratio can be estimated as following:
\be {(qq)_3 \over q}\approx {3+27 \over 12} exp({\mu-M+|\delta E|
\over T})2^{3/2} \ee where the binding $|\delta E|$ is negligible
(actually only the 3 $(qq)_3^{J=I=0}$ states are very likely
bound)\footnote{The last factor comes from $M^{3/2}$ in the
pre-exponent, originated from momentum integral.}. At small $\mu$
where the data under consideration are calculated, $M/T\approx 5$
and their contribution is at few percent level, negligible
compared to uncertainties.

{\bf Baryons:} as we found  in \cite{LS} they are bound till about $T=1.6T_c$.
In the 2-flavor theory they are
the $N,\Delta$ 3-quark states. Only the s-wave basic states
survive above $T_c$, while all other resonances (used in the first
section at $T<T_c$) which are orbital or radial excitations of
$N,\Delta$ families are ``melted''.

The baryons are also numerous (20) but the suppression factor due
to mass is much smaller \be {(qqq) \over q}\approx {20 \over 12}
exp({2\mu-2M+|\delta E| \over
 T})
3^{3/2}
 \ee
Near the ``endpoint'' of baryons with zero binding (which according
to \cite{LS} is at $T=1.6Tc$) their mass is $3M_q$, expected to be in the
range of 2.5- 3 GeV. As it is an order of magnitude larger than $T$, one
would not expected to contribute to pressure etc.

However, unlike the quark, quark-gluon and diquarks (which after
all are colored objects existing only above $T_c$), $N,\Delta$ baryons are
colorless and thus
survive on both sides of the boundary of (a continuous) phase
transition (a crossover, more accurately), thus the masses of
baryons at $T\rightarrow Tc$ are expected to join continuously to
their known values at lower $T$. This of course implies that the
binding energy near $T_c$ gets very large due to some deeper yet
poorly-known mechanism, and the potential model used in \cite{LS}
to evaluate this binding will not be applicable. The situation is
basically the same as with mesons: as emphasized in
\cite{SZ_bound} the pion mass must (by definition of chiral
breaking) vanish (in the chiral limit) at $T\rightarrow Tc$, which
potential model also cannot reproduce. we will use below
the following parameterization (in
$T_c$ units)
\begin{equation}
M_N = 9.5+4.6 * \tanh [3.8 * (T-1.4)]
\end{equation}
\begin{equation}
M_{\Delta} = 10.25+3.85 * \tanh [3.8 * (T-1.4)]
\end{equation}
interpolating between the nucleon and $\Delta$
vacuum masses at low $T$, while
 approaching the same value $3M_q$ at high temperatures.
We plot it in Fig.\ref{fig_mass}, together with
  the masses of various other states to be
used in later   in
Fig.\ref{fig_c4_baryons} for susceptibilities.
The main feature is fundamentally again enforcing confinement:
 when going from QGP  side toward $T_c$,
all colored degrees of freedom get extremely heavy and drop out
from system, while all colorless degrees of freedom
 get more tightly
bound and eventually dominate.

\begin{figure}
\hspace{+0cm} \centerline{\epsfxsize=7.cm\epsffile{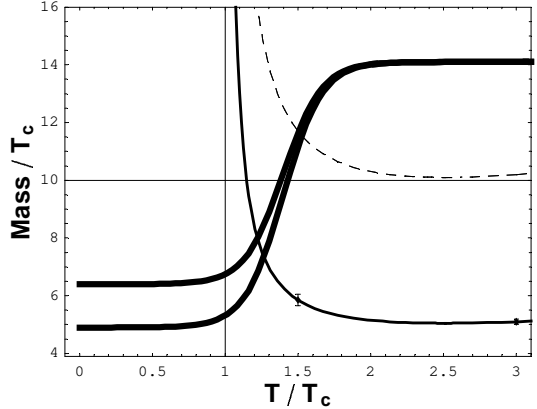}}
\hspace{+0.5cm}
 \caption{\label{fig_mass}
Masses of various states studied in this work. The thin solid line
is for quark and the dashed line is twice quark mass which is
roughly for quark-gluon and diquark. The lower thick solid line is
for nucleon states and the upper one for $\Delta$ states. These
masses are used for calculation of Fig.\ref{fig_c4_baryons}. }
 \end{figure}

  We now approach the central point of the paper:
the baryon contribution provides a natural interpretation of the
 structures observed in susceptibilities measured on the lattice, the
 large peak near $T_c$ in $d_4(T)$ and a more complicated
"wiggle" structure is seen in $d_6(T)$.
This happens because
the expected mass dependence of baryons on $T,\mu$, shown in
Fig.\ref{fig_mass}, should have a characteristic shape with an inflection
point, separating the region in which the second derivative $M_B''$
is negative (above $T_c$) and positive (below $T_c$). That is why
the contributions of the baryons
to $d_6$ show a ``wiggle''
 as seen from the corresponding curves in
Fig.\ref{fig_c4_baryons}. Note also, that there is a less pronounced
wiggle of the same origin in baryonic $d_4$: we think its negative
part is the reason why the $qg$ and $qq$ contributions above $T_c$
can get compensated and by coincidence the $d_4/d_2$ ratio gets
close to 1 there.

One additional argument for baryonic nature of the structures seen
in $d_4,d_6$ is the following one. Each derivative over $\mu_q$
leads to factor 3, so 2 of them give 9. If instead one has two
derivatives over $\mu_I$ the factor obtained is $(2I_3)^2$, which
is 1 for $p,n,\Delta^+,\Delta^0$ and 9 for $\Delta^{++},\Delta^-$.
As a result, if one ignores the mass difference between these
states, one finds that baryonic contribution to both should have
the ratio $d_n^I/d_n=(1/9)*(4/20)+(1+1/9)(8/20)=.467$, while this
ratio should be 1 for ideal quark gas. The actual ratio of these
quantities according to UKB data are shown in Fig.\ref{fig_ratio}.
We see near $T_c$ the data obviously favor the existence of
baryons, especially for $d_6^I/d_6$, and the quark asymptotic end
is arrived at about $1.4T_c$ for $d_4^I/d_4$ while only after
$1.8T_c$ for $d_6^I/d_6$. These evidences strengthen the necessity
 of  baryonic interpretation of the higher susceptibilities.
\begin{figure}
\hspace{+0cm} \centerline{\epsfxsize=6.cm\epsffile{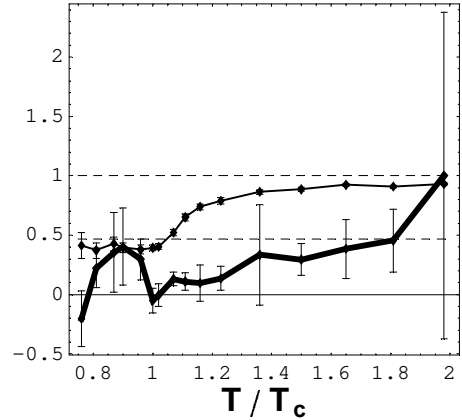}}
\hspace{+0cm}
 \caption{\label{fig_ratio}
The susceptibilities ratios $d_4^I/d_4$ (the thin solid) and
$d_6^I/d_6$ (the thick solid). The dashed lines correspond to
ideal quark gas (upper) and ideal baryonic gas (lower). }
 \end{figure}

{\bf Taking everything together}, including quarks, quark-gluons,
diquarks and baryons, we arrived at summary plots  shown in
Fig.\ref{fig_c4_baryons}. We repeat that
all masses used are as shown in Fig.\ref{fig_mass} and their
$\mu-$dependence is introduced in the same way according to
(\ref{mass_mu}).

( The
bound states' "endpoints" are set to be $2.1T_c$ for $(qg)_3$
quark-gluons, $1.4T_c$ for $(qg)_6$, $1.4T_c$ for diquarks, and
$1.6T_c$ for baryons, according to \cite{LS}. The gradual removal
near melting point is done by similar means as in \cite{SZ_bound}.
The results are shown in Fig.\ref{fig_c4_baryons}, where the
overall values as well as the contributions of each kind of states
are all present.)

Let's focus on the $T>T_c$ side. The conclusions are: \\(i) as
expected the diquark contribution is negligible for all three
quantities even after including the suspect $(qq)_3^{J=I=1}$
states, but it is clearly growing as getting to higher
derivatives;
\\(ii)For $d_2$ quark provides main contribution, and we emphasize
the fitting will be much better if we include the large binding of
$qg$ states near $T_c$. We have shown above that large uncertainty
in its binding, including polymers, would allow for a good fit
here, which we decided not to do.
\\(iii) In $d_4$ it is precisely the baryons that produce the
desired large peak near $T_c$ till about $1.3T_c$ where quarks
become important;
\\(iv) The baryons' contribution extremely dominant the behavior
of $d_6$, especially the "wiggle" shape.

We conclude that two prominent structures, a peak in $d_4(T)$ and
a ``wiggle'' in $ d_6(T)$ are naturally reproduced by baryons.

\begin{figure}
\hspace{+0cm} \centerline{\epsfxsize=7.cm\epsffile{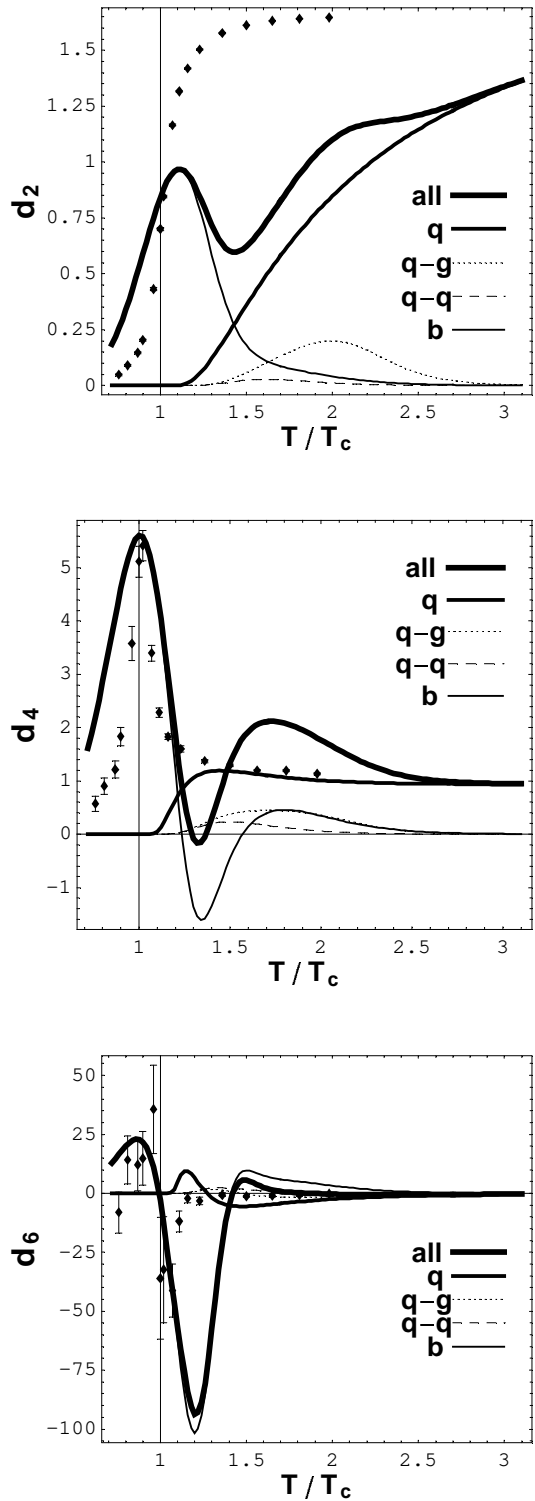}}
\hspace{+0.5cm}
 \caption{\label{fig_c4_baryons}
The contributions of different states to (a) $d_2$, (b) $d_4$ and
(c)  $d_6$, as well as the summed total values. The thickest solid
lines are for taking all together, while the medium solid lines
for quark, the thin solid lines for baryon, the dotted lines for
quark-gluon, and the dashed lines for diquark, respectively.}
 \end{figure}

\section{Summary}
 In one sentence, the main lessons from the UKB susceptibilities
is that the baryons $N\Delta$ do survive the
QCD phase transition, but are rapidly becoming quite heavy across it.

More generally, the discussed data set on the baryonic and isospin
susceptibilities at $T>T_c$ can be  described in two different
scenarios. (i) The first is a  quark quasiparticle gas, with the
effective mass which is strongly $decreasing$ near the phase
boundary into the QGP phase; (ii) the second is a picture
including baryons  with the mass rapidly $increasing$  across the
phase boundary toward QGP, to about $3M_q$.

 The first scenario was already pointed out by BKS \cite{BKS},
while our discussion makes it a bit more general.
Its attractive features notwithstanding,
it suggests  the
 values of the mass  not large enough to accommodate
the existing constraints from other lattice measurements.
We also think it is not possible to have quark degrees of freedom
in hadronic phase. Thus
 we conclude that success of such scenario is unlikely.

The second scenario, based on baryons,
 can  provide another explanation of the main features of the data,
namely the observed
peak in
 $d_4(T)$ and a ``wiggle'' in $ d_6(T)$. It also  naturally
 explains the flavor-changing $d_4^{ud}, d_6^{ud}$, which are not small
relative to flavor-diagonal ones. Last but not least,
this scenario provides
 a desired continuity to the baryon resonance
gas picture at $T<T_c$.

 Although the
susceptibilities $d_n(T)$ we used in this work are highly
 sensitive tools, they are quite indirect. Thermodynamical observable
in general cannot tell the difference between ``melting'' baryons
(getting unbound) and baryons remaining well bound but just
getting too heavy: in both cases all one finds is that
 their contribution to thermodynamics
effectively disappears.
 Besides, the
 ideal gas models used in these studies are probably too naive
to claim really quantitative description of the data. One should
instead study directly the spectral densities of the  correlators
of the appropriate baryonic currents ($qqq$) and see if there are
baryonic peaks there, like what has been done for charmonium and
light mesonic channels. Only such direct
measurements
would tell us  which scenario is the correct one.

Speaking about experimental confirmation of the ``bound state''
scenario, we think the best chance could be observation of the vector
mesons.
As described in detail in \cite{Casalderrey-Solana:2004dc}, vector
mesons
$\rho,\omega,\phi$ are expected to become heavy near their
disappearence point, like the baryons discussed above, reaching
 the mass $\approx 2M_q=1.5-2\, GeV$. The next
generation of RHIC dilepton experiments have a chance to see if this
is indeed what is happening in QGP.

{\bf Acknowledgments.\,\,} This work was partially supported by
the US-DOE grants DE-FG02-88ER40388 and DE-FG03-97ER4014. One of
us (ES) thanks K. Redlich for the discussion of these lattice
data, which triggered this work.


\end{narrowtext}
\end{document}